\newcounter{myctr}
\def\myitem{\refstepcounter{myctr}\bibfont\noindent\ifnum\themyctr>9\else\phantom{0}\fi\hangindent17pt\themyctr.\enskip}

\documentclass{ws-ijqi}

\begin{document}

\markboth{Jonas Maziero, Roberto M. Serra}
{Classicality witness for two-qubit states}

\catchline{}{}{}{}{}

\title{CLASSICALITY WITNESS FOR TWO-QUBIT STATES}

\author{JONAS MAZIERO}

\address{{Centro de Ci\^{e}ncias Naturais e Humanas, Universidade Federal do ABC, R.
Santa Ad\'{e}lia 166, 09210-170, Santo Andr\'{e}, S\~{a}o Paulo, Brazil}\\
jonas.maziero@ufabc.edu.br}

\author{ROBERTO M. SERRA}

\address{{Centro de Ci\^{e}ncias Naturais e Humanas, Universidade Federal do ABC, R.
Santa Ad\'{e}lia 166, 09210-170, Santo Andr\'{e}, S\~{a}o Paulo, Brazil}\\
serra@ufabc.edu.br}

\maketitle

\begin{history}
\received{Day Month Year}
\revised{Day Month Year}
\end{history}

\begin{abstract}
In the last few years one realized that if the state of a bipartite system can be written as $\sum_{i,j}p_{ij}|a_{i}\rangle \langle a_{i}|\otimes |b_{j}\rangle \langle b_{j}|$, where $\{|a_{i}\rangle \}$ and $\{|b_{j}\rangle \}$ form orthonormal basis for the subsystems and $\{p_{ij}\}$ is a probability distribution, then it possesses at most classical correlations. In this article we introduce a nonlinear witness providing a sufficient condition for classicality of correlations (absence of quantum discord) in a broad class of two-qubit systems. Such witness turns out to be necessary and sufficient condition in the case of Bell-diagonal states. We show that the witness introduced here can be readily experimentally implemented in nuclear magnetic resonance setups.
\end{abstract}



\section{Introduction}

The characterization and quantification of quantum and classical correlations presented in quantum systems are among the principal and more interesting problems in quantum information science (QIS). The seeds for this program can be ascribed to the papers of Einstein, Podolsky and Rosen\cite{EPR} and Schr\"{o}dinger\cite{Schrodinger}, who somehow introduced us to the notions of nonlocal correlations and non-separability in composed quantum systems. In this context, early discussions about quantum and classical correlations attributed the difference between the two types of correlations to the nonlocal character of the former, which was associated with the violation of Bell's inequalities\cite{Bell}. Subsequently, Werner gave an operational characterization of quantum correlations (at that time considered as synonymous of entanglement) as being the ones that cannot be generated by local operations and classical communication (LOCC)\cite{Werner}. The development of these ideas led to the today named theory of entanglement, that turned out to be a fruitful branch of research (see Ref. \refcite{Horodeckis} for a review). On the other side, recent studies have shown that entanglement is not the last word regarding the quantumness of correlations in composed quantum systems. Based on information-theoretical concepts, Ollivier and Zurek introduced the so called quantum discord as a measure of quantum correlations and showed that the correlations in a bipartite mixed state can have a quantum character even if it is separable\cite{Ollivier}. Oppenheim and co-workers came out with a similar conclusion from another quantum correlation measure, the quantum deficit, that was proposed based on a physical perspective\cite{Oppenheim}. After these early works on this subject, several measures of quantum correlation were proposed and analyzed (see Refs. \refcite{Maziero1} and \refcite{Celeri-Review} for a partial list of references). In particular the quantum discord has received a lot of attention. It has been subjected to experimental tests\cite{Lanyon,Xu,Diogo,Auccaise}, being recognized as a resource in several contexts\cite{Zurek,Datta,Lutz,Shabani,Cavalcanti}, showing peculiar dynamic behavior under decoherence\cite{Maziero3,Maziero4,Mazzola,Ferraro}, and exhibiting an interesting link to quantum phase transitions\cite{Sarandy,Chen,Maziero5,Maziero6}. Our goal in this work is not to make directly use of these more general measures of quantum correlations, but to introduce a witness for them.


\section{A Classification of Quantum States}

A brief summary concerning a classification of bipartite quantum states with relation to its correlations is in order. Any bipartite state that can be created via LOCC is said to be separable and its more general form reads
\begin{equation}
\sum_{i}p_{i}\rho _{i}^{a}\otimes \rho _{i}^{b},
\end{equation}
where $\{p_{i}\}$ is a probability distribution and $\rho _{i}^{a}$ and $\rho _{i}^{b}$ are quantum states for the two subsystems. By definition, a quantum system is entangled if its state is not separable.  But contrary to our naive intuition, separability and classicality of correlations are not the same issue. In fact separable states can also possess quantum characteristics in its correlations. Actually, a system is at most classically correlated only if its state can be written as
\begin{equation}
\sum_{i,j}p_{ij}|a_{i}\rangle \langle a_{i}|\otimes |b_{j}\rangle \langle b_{j}|,  \label{CC}
\end{equation}
with $\{|a_{i}\rangle \}$ and $\{|b_{j}\rangle \}$ forming orthonormal basis for the two subsystems and $\{p_{ij}\}$ being a probability distribution. Piani and colleagues gave an interesting characterization of this class of states as being the only ones whose correlations can be locally broadcast\cite{Piani}. 

The quantum correlations in bipartite states $\rho$ that cannot be cast as in Eq. (\ref{CC}) are quantified, for instance, by the so called \textit{quantum discord}, that can be defined as the difference between two classically-equivalent expressions for the mutual information:
\begin{equation}
\mathcal{D}(\rho)=\mathcal{I}(\rho)-\max_{\hat{O}}\mathcal{J}(\rho). \label{discord}
\end{equation}
The quantum mutual information,
\begin{equation}
\mathcal{I}(\rho)=S(\rho^{a})+S(\rho^{b})-S(\rho),
\end{equation} 
is a quantifier for the total (quantum plus classical) correlation between the subsystems $a$ and $b$, where $S(\rho^{x})=-\mathrm{Tr}(\rho^{x}\log_{2}\rho^{x})$ is the von Neumann entropy, which measures the uncertainty about the system $x$, and $\rho^{a(b)}=\mathrm{Tr}_{b(a)}(\rho)$. The other version for mutual information reads 
\begin{equation}
\mathcal{J}(\rho)=S(\rho^{a})-\sum_{j}\mathrm{Pr}(o_{j})S(\rho_{j}^{a}),
\end{equation}
and quantifies the information obtained about the subsystem $a$ when the observable $\hat{O}=\sum_{j}o_{j}|o_{j}\rangle\langle o_{j}|$ is measured on subsystem $b$. The maximization in Eq. (\ref{discord}) is intended to use the observable $\hat{O}$ whose measurement yields the maximal amount of information about the subsystem $a$'s state. The state of the subsystem $a$ after the result $o_{j}$ is obtained, with probability
\begin{equation}
 \mathrm{Pr}(o_{j})=\mathrm{Tr}(\mathbf{I}^{a}\otimes|o_{j}\rangle\langle o_{j}|\rho),
\end{equation}
in the measurement of $\hat{O}$ is given by
\begin{equation}
 \rho_{j}^{a}=\frac{1}{\mathrm{Pr}(o_{j})}\mathrm{Tr}_{b}[(\mathbf{I}^{a}\otimes|o_{j}\rangle\langle o_{j}|)\rho(\mathbf{I}^{a}\otimes|o_{j}\rangle\langle o_{j}|)],
\end{equation}
where $\mathbf{I}^{ab}$ is the identity operator in the Hilbert's space $\mathcal{H}_{ab}$.

At last in our state classification comes the product states, 
\begin{equation}
\rho^{a}\otimes\rho^{b},
\end{equation}
which possesses no correlations at all, where $\rho ^{a}$ and $\rho ^{b}$ are density operators of the subsystems.

A typical problem in QIS is to quantify how far a given state and the aforementioned states are one from the another or simply to distinguish them. The former task is ordinarily performed using measures of correlation, that are ultimately obtained from experimental state tomography. But in some circumstances it is enough to know, for example, if the correlations in the system are classical or quantum. In these situations we would like to witness classicality, in analogy to what is done with entanglement witnesses\cite{Terhal}, without doing the usually demanding quantum state tomography and also avoiding the generally hard numerical optimization procedures needed for the calculation of measures of quantum correlation. However, in contrast to the space of separable states, the set of classically correlated states is not convex. For that reason, as was proved by Rahimi and SaiToh\cite{Rahimi1}, a linear witness cannot do the job in general. In Ref. \refcite{Rahimi1}, a nonlinear quantum correlation witness, whose calculation involves a maximization over the set of classical correlated states, was proposed and computed for some very specific cases. In this article we introduce a nonlinear witness providing a sufficient condition for the classicality of correlations in a wide class of two-qubit states. For Bell-diagonal states, such witness is necessary and sufficient condition for the absence of quantumness in the correlations of the system. As will be shown in the sequence, the classicality witness introduced here can be readily implemented in experimental contexts such as, for example, in nuclear magnetic resonance (NMR) setups.


\section{Witness for Quantum Correlations}

Here we are interested in systems whose state takes the following form:
\begin{equation}
\rho =\frac{1}{4}\left(\mathbf{I}^{ab}+\vec{x}\ldotp\vec{\sigma}^{a}\otimes \mathbf{I}^{b}+\mathbf{I}^{a}\otimes \vec{y}\ldotp\vec{\sigma}^{b}+\sum_{i=1}^{3}c_{i}\sigma _{i}^{a}\otimes\sigma _{i}^{b}\right), \label{2qubit}
\end{equation}
where $c_{i}\in \Re $, $\vec{x},\vec{y}\in \Re ^{3}$ are constrained such that the eigenvalues of $\rho $ are not negative. Besides $\mathbf{I}^{k}$ is the identity operator acting on state space of system $k=a,b,ab$ and $\vec{\sigma}^{j}=(\sigma _{1}^{j},\sigma _{2}^{j},\sigma _{3}^{j})$ with $j=a,b$, where $\sigma _{1}^{j}=|0\rangle \langle 1|+|1\rangle \langle 0|$, $\sigma _{2}^{j}=-i(|0\rangle \langle 1|-|1\rangle \langle 0|)$, and $\sigma_{3}^{j}=|0\rangle \langle 0|-|1\rangle \langle 1|$ are the Pauli operators
acting in the state space of the subsystem $j$ and $\{|0\rangle ,|1\rangle \}$ is the usual computational basis. It is worthwhile to mention that this class of states is quite general and it appears routinely in several theoretical\cite{Maziero3,Maziero4,Sarandy} 
and experimental (as, for example, in optical and NMR setups\cite{Xu,Diogo}) contexts.

Let us regard observables represented by the following set of hermitian operators: 
\begin{eqnarray}
\hat{O}_{i}&=&\sigma_{i}^{a}\otimes\sigma_{i}^{b}, \\
\hat{O}_{4}&=&\vec{z}\ldotp\vec{\sigma}^{a}\otimes\mathbf{I}^{b}+\mathbf{I}
^{a}\otimes\vec{w}\ldotp\vec{\sigma}^{b},
\end{eqnarray}
where $i=1,2,3$ and $\vec{z},\vec{w}\in\Re^{3}$ with $||\vec{z}||=||\vec{w}||=1$. We observe that the directions $\vec{z}$ and $\vec{w}$ should be picked out randomly. Now we consider a relation among these observables as follows 
\begin{equation}
W_{\rho}=\sum_{i=1}^{3}\sum_{j=i+1}^{4} |\langle \hat{O}_{i}\rangle_{\rho}\langle 
\hat{O}_{j}\rangle_{\rho}|,  \label{witness}
\end{equation}
where $\langle \hat{O}_{i}\rangle_{\rho}=\mathrm{Tr}(\hat{O}_{i}\rho)$ and $|x|$ is the absolute value of $x$. We see that $W_{\rho}=0$ if and only if the average value of at least three of the four observables defined above is zero. Thus, if we note that $\langle\hat{O}_{i}\rangle_{\rho}=c_i$ for $i=1,2,3$ and $\langle\hat{O}_{4}\rangle_{\rho}=\vec{z}.\vec{x}+\vec{w}.\vec{y}$, it
follows that the only way in which we warrant that $W_{\rho}=0$ (independently of the directions $\vec{z}$ and $\vec{w}$) is if the state $\rho$ assumes the form of one of the following states 
\begin{eqnarray}
\chi_{i}&=&\frac{1}{4}\left(\mathbf{I}^{ab}+c_{i}\sigma_{i}^{a}\otimes\sigma_{i}^{b}\right), \\
\chi_{4}&=&\frac{1}{4}\left(\mathbf{I}^{ab}+ \vec{x}\ldotp\vec{\sigma}^{a}\otimes\mathbf{I}^{b}+\mathbf{I}^{a}\otimes\vec{y}\ldotp\vec{\sigma}^{b}\right),
\end{eqnarray}
where $i=1,2,3$. It turns out that all these four states can be straightforwardly set in the form of Eq. (\ref{CC}), and hence are
at most classically correlated. Therefore $W_{\rho}=0$ is a sufficient condition for $\rho$ to be classically correlated. Moreover, for the so called Bell-diagonal class of states, 
\begin{equation}
\rho^{bd}=\frac{1}{4}\left(\mathbf{I}^{ab}+\sum_{i=1}^{3}c_{i}\sigma_{i}^{a}\otimes\sigma_{i}^{b}\right),  \label{BD}
\end{equation}
$W_{\rho^{bd}}=0$ is a necessary and sufficient condition for classicality. This result follows by noting that, in this case, $\rho^{bd}$ being classical correlated implies that it must take the form:
\begin{equation} 
\frac{1}{4}\left(\mathbf{I}^{ab}+c_{i}\sigma_{i}^{a}\otimes\sigma_{i}^{b}\right),
\end{equation} 
with $i=1$ or $i=2$ or $i=3$ (see e.g. Ref. \refcite{Lang}), and thus implies $W_{\rho^{bd}}=0$.

\subsection{Example: Correlations in the Werner's State}

As an example we apply the witness given in Eq. (\ref{witness}) to the Werner's state, 
\begin{equation}
\rho^{w}=(1-\alpha)\frac{\mathbf{I}^{ab}}{4}+\alpha|\Psi^{-}\rangle\langle\Psi^{-}|,
\label{werner}
\end{equation}
where $0\le\alpha\le1$ and
\begin{equation}
|\Psi^{-}\rangle=\frac{1}{\sqrt{2}}(|01\rangle-|10\rangle).
\end{equation}
By completeness we recall that the Werner's state violates the CHSH inequality\cite{CHSH} for $\alpha\ge1/2$ and violates the Peres-Horodecki criterion\cite{Peres,Horodeckis1} for $\alpha>1/3$. By a direct calculation one obtain that, for this state, $W_{\rho^{w}}=3\alpha^{2}$. As the Werner's state belongs to the Bell-diagonal class, and in this case $W_{\rho^{w}}=0$ is necessary and sufficient condition for classicality, it follows that $\rho^{w}$ possesses quantumness in its correlations for all $\alpha\neq0$. It is worth mentioning that the same result is obtained when we use the quantum discord to study the character of correlations in the state (\ref{werner})\cite{Ollivier}.


\section{Experimental Implementation}

In what follows we present some relations between correlation functions and magnetizations showing that the classicality
witness introduced in this article can be readily implemented using the already developed tools of nuclear magnetic resonance (NMR). In these systems the qubits are encoded using nuclear spins and unitary transformations are obtained through suitable sequences of radio-frequency pulses. The natural observables in NMR experiments are the local transverse magnetizations, which are obtained directly from the NMR signal\cite{NMR}. Let us consider the following equalities
\begin{eqnarray}
&&\sigma_{1}^{a}\otimes\sigma_{1}^{b}=\mathrm{CNOT}_{a\rightarrow b}(\sigma_{1}^{a}\otimes\mathbf{I}^{b})\mathrm{CNOT}_{a\rightarrow b}, \\ 
&&\sigma _{2}^{a}\otimes \sigma _{2}^{b} =R_{3}^{\dagger }\left( \sigma _{1}^{a}\otimes\sigma _{1}^{b}\right) R_{3}, \\ 
&&\sigma _{3}^{a}\otimes \sigma _{3}^{b}=R_{2}^{\dagger }\left( \sigma _{1}^{a}\otimes\sigma _{1}^{b}\right) R_{2},
\end{eqnarray}
where 
\begin{eqnarray}
&&\mathrm{CNOT}_{a\rightarrow b} =|0\rangle \langle 0|\otimes \mathbf{I}^{b}+|1\rangle \langle 1|\otimes \sigma _{1}^{b}, \\ 
&&R_{k} =R_{k}^{a}(\pi /2)\otimes R_{k}^{b}(\pi /2), \\
&&R_{k}^{j}(\pi/2 )=\cos(\pi/4)\mathbf{I}^{j}-i\sin(\pi/4)\sigma_{k}^{j},
\end{eqnarray}
with $j=a,b$ and $k=2,3$. Now, if we define the states 
\begin{eqnarray}
\eta  &=&\mathrm{CNOT}_{a\rightarrow b}(\rho )\mathrm{CNOT}_{a\rightarrow b}, \label{eta} \\
\zeta  &=&\mathrm{CNOT}_{a\rightarrow b}(R_{3}\rho R_{3}^{\dagger})\mathrm{CNOT}_{a\rightarrow b}, \label{zeta} \\
\xi  &=&\mathrm{CNOT}_{a\rightarrow b}(R_{2}\rho R_{2}^{\dagger})\mathrm{CNOT}_{a\rightarrow b}, \label{xi}
\end{eqnarray}
then the following set of relations between correlation functions and 
magnetizations is obtained 
\begin{eqnarray}
\langle \sigma _{1}^{a}\otimes \sigma _{1}^{b}\rangle _{\rho } &=&\langle
\sigma _{1}^{a}\otimes \mathbf{I}^{b}\rangle _{\eta }, \\
\langle \sigma _{2}^{a}\otimes \sigma _{2}^{b}\rangle _{\rho } &=&\langle
\sigma _{1}^{a}\otimes \mathbf{I}^{b}\rangle _{\zeta }, \\
\langle \sigma _{3}^{a}\otimes \sigma _{3}^{b}\rangle _{\rho } &=&\langle
\sigma _{1}^{a}\otimes \mathbf{I}^{b}\rangle _{\xi}.
\end{eqnarray}
Thus, by looking at these relations, one can note that the classicality witness defined in Eq. (\ref{witness}) can be straightforwardly implemented in NMR setups\cite{NMR}. More specifically, the correlation functions $\langle\sigma_{i}^{a}\otimes\sigma_{i}^{b}\rangle_{\rho}$ are obtained by running the experiment three times. In each realization of the experiment, one must prepare the system in the state $\rho$ and, after doing the local-unitary and controlled-NOT operations to achieve the states $\eta$, $\zeta$, and $\xi$ (as shown in Eqs. (\ref{eta})-(\ref{xi})), measure the magnetization in the $x$-direction on qubit $a$.


\section{Concluding Remarks}

It is important to stress that a crucial aspect that one should take into account when dealing with witnesses (or criteria)  for the presence (or absence) of quantumness in the correlations of a composed system (for recent related works see Refs. \refcite{Rahimi1,Bylicka,Dakic}, and \refcite{Chen1}) is not only to reduce the number of required experimental settings in relation to those involved in quantum state tomography, but also to escape the hard numerical optimization procedures generally involved in the evaluation of quantum correlation measures. Here we have introduced a nonlinear witness providing a sufficient condition for classicality in the correlations of a broad class of two-qubit systems. Such a witness is necessary and sufficient condition in the case of Bell-diagonal states. The classicality witness introduced in this work has a straightforward experimental implementation, 
precluding any additional numerical optimization process\cite{Rahimi1} or ancillary qubits\cite{Zhang}, as required by other proposals. Such feature is an important advantage for experimental bench tests of classicality. Furthermore, by regarding some useful relations between correlation functions and magnetizations we observed that the classicality witness given in Eq. (\ref{witness}) can be readily experimentally implemented  in, for instance, NMR setups.

\section*{Acknowledgments}

We are grateful for the funding from UFABC, CAPES, FAPESP, and the Brazilian National Institute for Science and Technology of Quantum Information (INCT-IQ). We thank L. C. C\'eleri for discussions.


\end{document}